\documentclass[showpacs,preprintnumbers,amsmath,amssymb,twocolumn,floatfix]{revtex4}
\usepackage[usenames]{color}
\usepackage{epsfig}
\usepackage{amssymb}

\newcommand{\be}{\begin{equation}}
\newcommand{\ee}{\end{equation}}
\newcommand{\bee}{\begin{eqnarray}}
\newcommand{\eee}{\end{eqnarray}}

\newcommand{\piN}{$\pi N \:$}
\newcommand{\etaN}{$\eta N \: $}

\newcommand{\piNetaN}{$\pi N \! \rightarrow \! \eta N \:$}

\definecolor{grey}{rgb}{0.9,0.9,0.9}
\definecolor{black}{rgb}{0,0,0}

\def \black{\color{black} \bf}

\begin{document}

\title{ The \piNetaN data demand the existence of N(1710)~P$_{11}$ resonance \\  reducing the 1700 MeV continuum ambiguity}

\author{S. Ceci, A. \v{S}varc and B. Zauner}

\address{Rudjer Bo\v{s}kovi\'{c} Institute, \\
Bijeni\v{c}ka c. 54, \\ 
10 000 Zagreb, Croatia\\ 
E-mail: alfred.svarc@irb.hr}
\date{\today}

\begin{abstract}
In spite of prolonged polemics, the agreement on the existence of N(1710) P$_{11}$ resonance has not until now been reached, and the Particle Data Group declares it as a 3-star resonance only. We show that the proper inclusion of inelastic channels in the coupled-channel formalism indisputably demands the existence of N(1710) P$_{11}$ state, and that it presumably stays "hidden" within the continuum ambiguity of any typical single channel partial wave analyses. Consequently, its Particle Data Group confidence rating should be raised to a 4-star resonance.
\end{abstract} 

\maketitle
A central task of baryon spectroscopy is to establish a connection between resonant states predicted by various low energy QCD models and hadron scattering observables.
A reasonable way to proceed seems to be to identify poles of analytic scattering amplitudes which simultaneously describe all experimental data in all attainable channels with theoretically predicted resonant states. The following step is to uniquely extract resonant parameters out of obtained poles. The "missing resonance problem", a failure to experimentally confirm a number of predicted quark model states (standard, glue enriched or created from color neutral di- or multi-quark molecules)  \cite{Cap00}, poses a dilemma whether to suspect the reliability of quark models or to mistrust the resonance identification in partial-wave analyses.

Having in mind a possible number of new states to be identified, a dispute about the existence of already reported resonant states, seen by one group and not confirmed by another one, presents the matter which should be cleared up with forceful efficiency. We concentrate on the N(1710) P$_{11}$ problem.

One of the latest and widely accepted partial wave analysis, Virginia Polytechnic Institute/George Washington University (GWU/VPI) \cite{Arn04}, does not see the N(1710) P$_{11}$ state. When the energy-dependent coupled-channel Chew-Mandelstam K matrix formalism is applied to fit $\pi N$ elastic and \piNetaN experimental data, the obtained FA02 (SP06)  solution contains no poles in the P$_{11}$ partial wave in the vicinity of 1710 MeV. On the other hand, the error bars of single energy solutions (SES) accompanying it on the web page \cite{GWUWEB} do show disproportionable  increase in that energy domain. As this analysis heavily relies on the elastic channel, the immanent continuum ambiguity problem \cite{Atk73,Bow75} is  handled by iteratively stabilizing the solution  during the minimization procedure through applying the fixed t dispersion relations. Yet, the suspicious error bars increase in SES has not been understood and the N(1710)~P$_{11}$ pole has not been established.

Other partial wave analyses do report the presence of the N(1710)~P$_{11}$ pole. Due to the constraints coming from the crossed channels, a single channel and fully analytic PWA (Karlsruhe-Helsinki KH80 \cite{KH84}) undoubtedly sees the N(1710)~P$_{11}$ state, and qualifies it as being strongly inelastic. Including inelastic channels leaves very little doubt about the N(1710)~P$_{11}$ existence. The coupled-channel T-matrix Carnegie-Mellon Berkeley (CMB) type models \cite{Cut79,Bat98,Vra00}, the coupled channel K-matrix analyses/University of Giessen \cite{Pen02}, coupled channel K-matrix analyses dealing with kaon-hyperon interactions \cite{Wen04} and the Kent state analysis \cite{Man92} are all very affirmative about the existence of the N(1710) P$_{11}$ resonance. The consequence of the current failure to achieve the unanimous agreement is that only a 3-star confidence rating is attributed to the N(1710)~P$_{11}$ resonance by the Particle Data Group \cite{PDG}. This rating should be changed.

In this Letter we present the essence of the mechanism how inelastic channels enforce the existence of the N(1710) P$_{11}$ state. Using the T-matrix coupled channel formalism of the CMB type \cite{Cut79} we show that fitting only elastic channel requires just the N(1440)~P$_{11}$ state, and  P$_{11}$ T matrix is smooth in the 1700 MeV range. When inelastic channels are also fitted, no doubt is left about the existence of the  N(1710)~P$_{11}$, but the smoothness of the  P$_{11}$ T matrix is spoiled. This finding contradicts the smooth energy behavior of the offered VPI/GWU FA02 (SP06) solution, but \emph{is not} in controversy with the collection of \emph{single energy solutions (SES)} of the same group.  We believe that reported enlargement of error bars of VPI/GWU SES in the critical energy range is not only of experimental, but primarily of theoretical nature, and we offer the explanation that a  N(1710) P$_{11}$ resonance is "hidden" within the continuum ambiguity  \cite{Atk73,Bow75} in the 1700 MeV range. The way how the \piNetaN data have been used in the VPI/GWU analysis is also not being of much help for eliminating the 1700 MeV continuum ambiguity: all data above T$_{\rm lab}$ = 800 MeV, important to see the N(1710) $P_{11}$ resonance, have unfortunately not been included in this analysis. So, no wonder the N(1710) P$_{11}$ is not seen in FA02 (SP06).

\begin{table*}[!ht]
\caption{The extracted P$_{11}$ partial wave T-matrix \vspace*{0.2cm} poles.} 
\begin{tabular}{c c | c c c c | c c c c | c c c c c}
 \hline \hline
\multicolumn{2}{c |}{} & \multicolumn{4}{c |}{$\pi N$ only / MeV} & \multicolumn{4}{c|}{$\eta N$ only / MeV}& \multicolumn{5}{c }{$\pi N$ and $\eta N$ / MeV} 
\\[+1ex]\cline{3-15} 
\multicolumn{2}{c |}{} & \multicolumn{4}{c |}{} & \multicolumn{4}{c|}{}& \multicolumn{5}{c }{} \\[-2ex]
$sol$ & $ N_{P} (N_{R})$&$\chi^2_{R}$& $ \textstyle \binom{\rm Re\, W}{\rm -2\, Im\, W}$ & $ \textstyle \binom{\ldots}{\ldots}$ & $\textstyle \binom{\ldots}{\ldots}$ & $\chi^2_{R}$
& $ \textstyle \binom{\rm Re\, W}{\rm -2\, Im\, W}$ & $\textstyle \binom{\ldots}{\ldots}$ & $\textstyle \binom{\ldots}{\ldots}$ & $\chi^2_{R}$ & $ \textstyle \binom{\rm Re\, W}{\rm -2\, Im\, W}$ & $\textstyle \binom{\ldots}{\ldots}$ & $\textstyle \binom{\ldots}{\ldots}$ & $\textstyle \binom{\ldots}{\ldots}$\\ [1.5ex]
\hline \hline
{}& {}& {}&  {}& {}& {}& {}& {}& {}& {}& {}& {}& {}& {} & {} \\[-2.5ex]
${}$ 1/4/7 & 3 (1)& 3.37 & $\textstyle \binom{1325}{ 175}$ & - & - &
1.24 & $ \textstyle \binom{1170}{75}$ & \colorbox{grey}{\black $ \textstyle { \binom{1735}{180}}$} & $ \textstyle
\binom{2175}{215}$ & 8.2 & $\textstyle \binom{1345}{150}$
 & \colorbox{grey}{\black$\textstyle {\binom{1880}{375}}$} & - & -\\
 {}& {}& {}& {}& {}& {}& {}& {}& {}& {}& {}& {}& {}& {} & {}\\[-2.5ex]
${}$ 2/5/8 & 4 (2) & 3.08 & $ \textstyle \binom{1335}{155}$ & $ \textstyle
\binom{1820}{290}$ & $ \textstyle
\binom{1970}{105}$ & 1.28 & $ \textstyle \binom{1405}{200}$ &
 \colorbox{grey}{\black $ \textstyle {\binom{1730}{170}}$} & $ \textstyle \binom{2175}{290}$ & 3.48 & $ \textstyle \binom{1350}{170}$ & \colorbox{grey}{\black $ \textstyle {\binom{1710}{80}}$} & $ \textstyle
\binom{1970}{330}$ & -\\ 
 {}& {}& {}& {}& {}& {}& {}& {}& {}& {}& {}& {}& {}& {}& {}\\[-2.5ex]
$ {}$ 3/6/9 & 5 (3) & 2.94 & $ \textstyle \binom{1320}{150}$ & $ \textstyle
\binom{1945}{140}$ & $ \textstyle \binom{1975}{ 190}$ & 1.36 & $ \textstyle \binom{1350}{195}$& 
\colorbox{grey}{\black $\textstyle {\binom{1730}{170}}$} & $ \textstyle \binom{2150}{310}$ & 2.71 & $ \textstyle \binom{1350}{170}$ & \colorbox{grey}{\black $\textstyle {\binom{1640}{330}}$} & \colorbox{grey}{\black $ \textstyle {\binom{1730}{150}}$} & $ \textstyle \binom{2120}{400}$\\ 
 {}& {}& {}& {}& {}& {}& {}& {}& {}& {}& {}& {}& {}& {} & {} \\[-2.5ex]
$ {} \ 10$ & 6 (4) & - & - & - & - & - & - & - & - & 2.86 & $ \textstyle
\binom{1350}{190}$ & \colorbox{grey}{\black $ \textstyle {\binom{1730}{215}}$} & \colorbox{grey}{\black $ \textstyle {\binom{1760}{260}}$} & $ \textstyle
\binom{2175}{170}$ \\ [1.5ex]
\hline\hline
\end{tabular}
\label{table1} 
\end{table*}  

Finally, let us just mention that the P$_{11}$ T matrix energy behavior obtained in this publication is almost identical to the (C-p-$\pi$+) solution  reported  by the Giessen coupled channel K-matrix analysis \cite{Pen02}. 
 
 \noindent \\
\textbf{\emph{Setting up the model}} 

 \indent
We use CMB model \cite{Cut79,Bat98,Vra00}, the separable coupled-channel partial-wave analysis with channel propagator $\phi$ analyticity ensured by once subtracted dispersion relation. The Bethe-Salpeter equation for dressed resonance propagator $G$ is explicitly solved by using bare propagators $G^0$ and the self energy $\Sigma$. The parameters of the model are bare propagator poles $s_0$, their number $N_P$, and channel-to-resonance mixing real matrix $\gamma$. Once the number of channels $N_C$, and bare propagator poles is chosen, the model contains total of $N_{P} + (N_{P} \times N_{C})$ free parameters per partial wave.

Our model contains resonant contributions and nonresonant background. The resonant contribution is generated by "dressing" the bare poles of the resonant propagator with the self energy term. Background is generated by two nonphysical poles lying beyond the measurable energy range. The initial number of genuine resonances, two less than N$_{P}$, is denoted as N$_{R}$.

The self energy is parameterized as $\Sigma=\gamma^T\Phi\gamma$, so the model will satisfy the unitarity demands. $\Phi$ is a diagonal matrix of channel propagators $\phi$, and Im$\phi$ are defined as in Ref. \cite{Bat98} for each channel. Unitary normalized T-matrix in the physical limit is given by $T = \sqrt{\mathrm{Im}\Phi} \, \gamma \,G \, \gamma^T \sqrt{\mathrm{Im}\Phi}$.

Each channel opening introduces additional Riemann sheet, every resonance state corresponds to a pole on each of them, and the most influential one lies on the nonphysical sheet closest to the physical axis. T-matrix poles are obtained as poles of dressed particle propagator $G$, and are reached by analytic continuation of channel propagators $\phi$ into the complex plane.

We use the model with three channels: two physical two body channels $\pi N$ and $\eta N$, and the third (effective) channel $\pi^2 N$, which is representing \emph{all remaining two and three body} processes in a form of a  two  body  process  with $\pi^2$ being a quasiparticle with a different mass  chosen  for each  partial  wave.

 \noindent \\
\textbf{\emph{The data base}} 

Model parameters are in principle obtained by fitting experimental data directly. However, as the formalism separates individual partial waves, we fit already available $\pi N$ elastic and \piNetaN partial-wave amplitudes, which we understand only as an effective representation of all existing measurements.

For $\pi N$ elastic partial-waves we have used a P$_{11}$ GWU single energy solutions \cite{GWUWEB,Arn04} having 122 data points with error bars.

Similarly as in ref. \cite{Kis04}, where the Pittsburgh results of the coupled channel PWA \cite{Vra00} are used as the experimentally constrained  T$^{{\rm S}_{11}}_{\pi N \,\eta N}$, we use the coupled channel model of the same type from Batini\'{c} et al  \cite{Bat98a} to obtain P$_{11}$ T matrix in \etaN channel. However, instead of using smooth theoretical curves, we construct 78 "experimental" data points by normally distributing the model input in order to simulate the statistical nature of really measured data.  The standard deviation $\sigma$ is set to 0.02, and that is almost exactly the average error value of GWU data. By using this procedure we were able to produce \piNetaN partial-wave data set that, when fitted, gives realistic $\chi^2$ values comparable to those obtained by SES fits.

Let us point out that both PWA \cite{Bat98a,Vra00} in addition to reproducing the standard set of \piNetaN data \cite{DATA} nicely reproduce the new measurements coming from the Crystal Ball Collaboration \cite{CRBALL}.
\\ \\
\noindent
\textbf{\emph{The fitting procedure}} \\ \indent
We start with a minimal set of resonances and increase their number until the satisfactory fit is achieved. We fit each channel separately and obtain individual collections of poles. Then we compare them. If collections disagree (different poles describe different channels), we conclude that various channels exhibit more sensitivity to particular resonances. As all of them have to exist they all have to be taken into consideration. In the end, we combine both channels and fit them  simultaneously with all poles established in individual fits regardless of the fact that extra poles might spoil the level of agreement in individual channels.  We increase the number of poles until the quality of fit, measured by the the lowest reduced $\chi^2$ value, can not be improved. In addition, a visual resemblance of the fitting curve to the data set as a whole is used as a rule of thumb; i.e. we are rejecting those solutions which have a tendency  to accommodate for the rapidly varying data points regardless of the $\chi^2$ value.
\\ \\
\textbf{\emph{Results and conclusions} } \\   \indent
For the total of 10 solutions, pole positions and the reduced $\chi^2_R$ (defined as $\chi^2$ divided by the difference of number of data points and number of fitting parameters) are given in Table 1.
\\ \\  \noindent
\emph{Elastic channel only \dotfill NO N(1710) P$_{11}$ } \\ \indent
We have fitted elastic channel only and obtained \mbox{$sols$~1-3}.
\begin{figure}[!h]
\hspace{-2.5cm}
\includegraphics[width=5.cm]{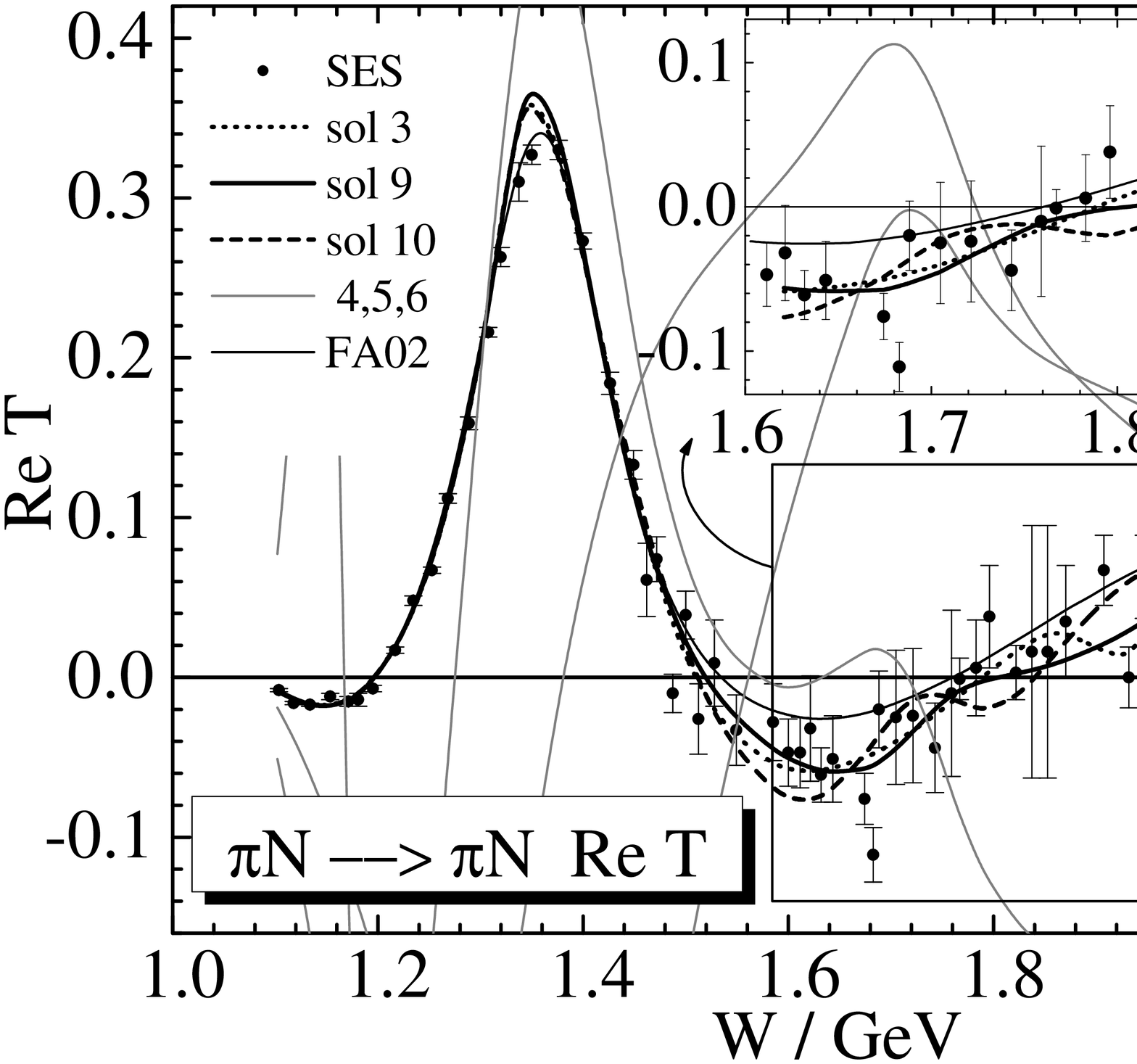}\\
 \hspace{-2.5cm}
\includegraphics[width=5.cm]{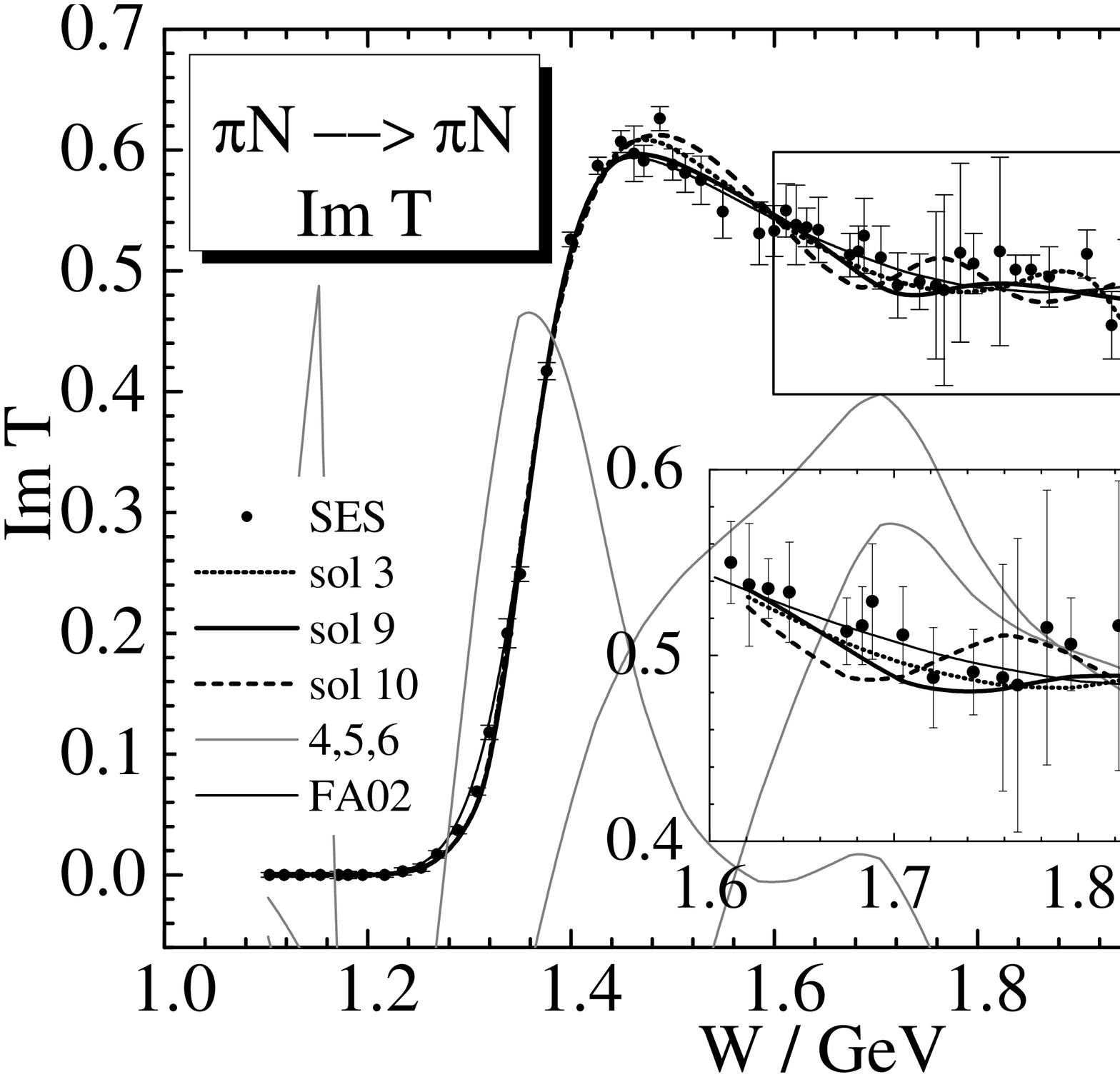}
 \caption{The agreement of the input $T_{\pi N \, \pi N}$ values with the results of the fit.}
\label{figure1}
\end{figure}
\\  \indent To achieve the overall agreement of the model with the experimental input of ref. \cite{GWUWEB,Arn04} only one physical pole in the vicinity of 1400 MeV (Roper) suffices. Adding new poles is just visually improving the quality of the high energy end of the fit, the $\chi^2_R$ is only slightly improved, changes are of cosmetic nature only, so the existence of the second pole near 1950 MeV is not essential but only consistent with the data.

The agreement of the input $T_{\pi N \, \pi N}$  values with the results of the elastic channel fit for the three-pole solution ($sol$~3) is given with dotted line in Fig.1, the predictions of $sols$~1 and 2 are not shown because they are very similar to $sol$~3.

The T-matrix values, predicted for the \mbox{\piNetaN} channel when only $\pi N$ elastic channel is fitted, strongly deviate from the \etaN input, and are just as an indication given with unlabeled thin grey lines in Fig.2.
\\ \\
\noindent
\emph{Inelastic channel only \dotfill N(1710) P$_{11}$ RECQUIRED} \\ \indent
We have fitted inelastic channel only and obtained $sols$~4-6.
\begin{figure}[!h]
\hspace{-2.5cm}
\includegraphics[width=5.cm]{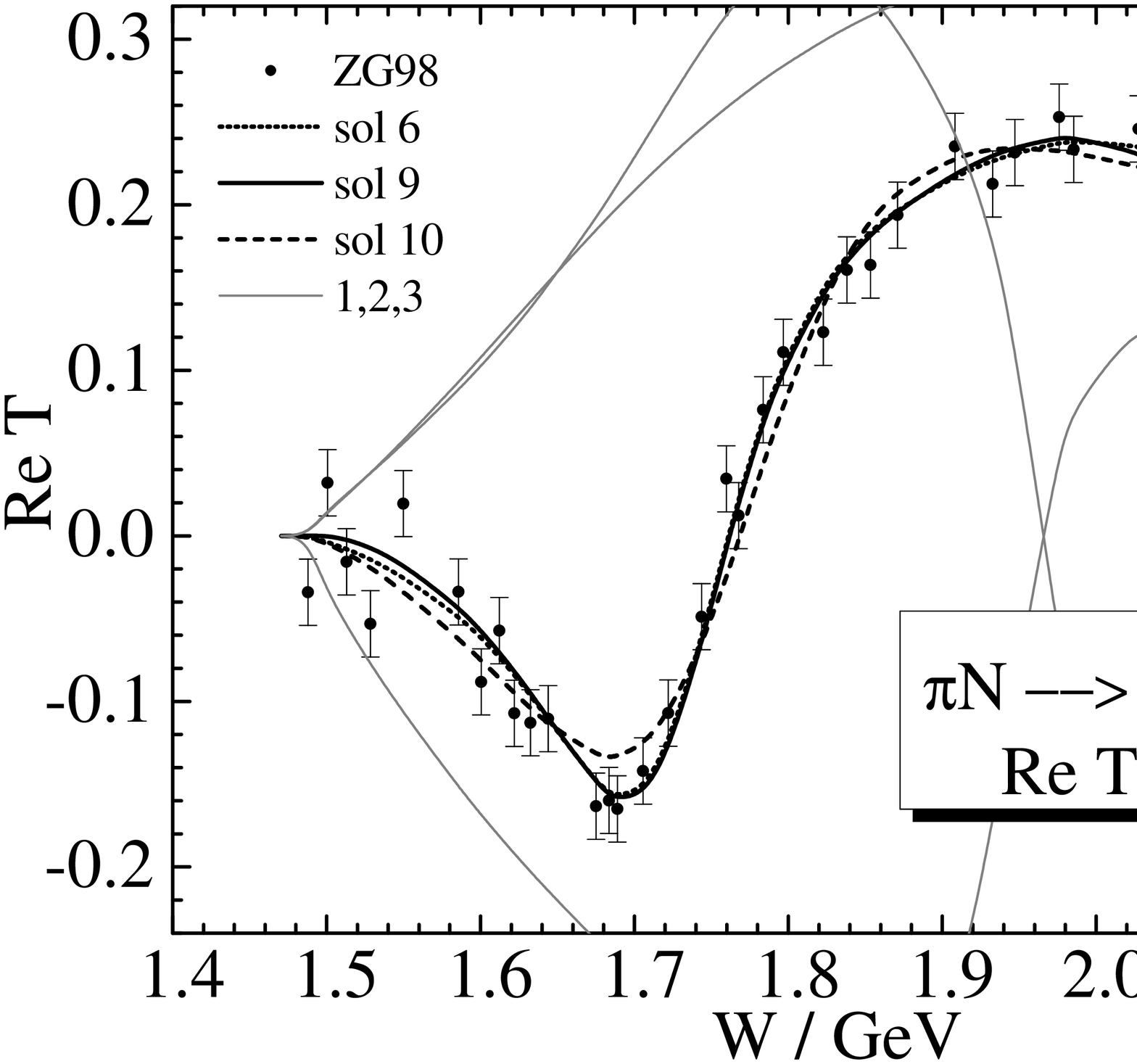}\\
 \hspace{-2.5cm}
\includegraphics[width=5.cm]{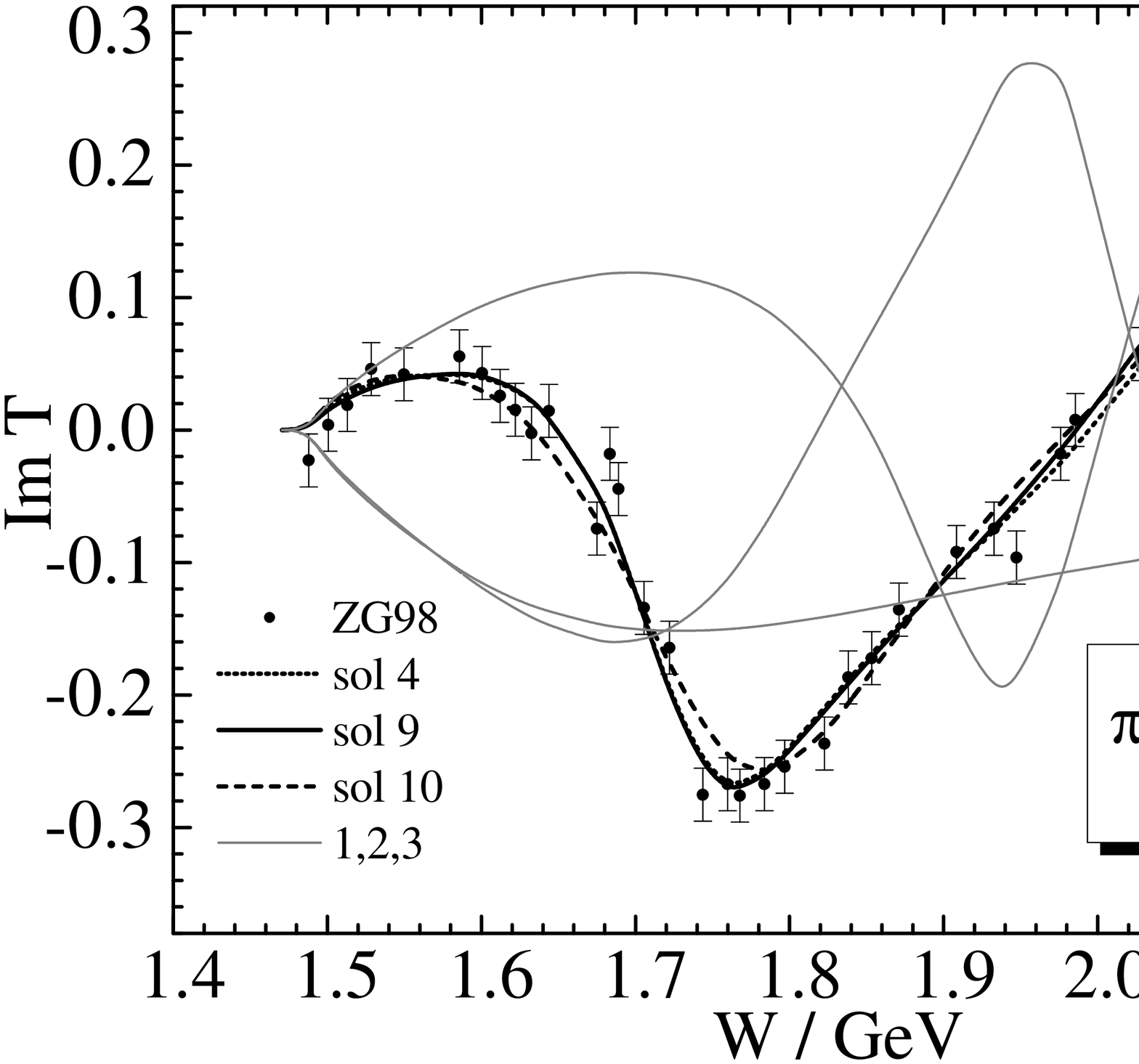}
 \caption{The agreement of the input $T_{\pi N \, \eta N}$ values with the results of the fit.}
\label{figure2}
\end{figure}
\\ \indent To achieve the overall agreement of the model with the experimental input of ref. \cite{Bat98a} we need at least three physical poles; one in the vicinity of 1300 MeV (below Roper), second one around 1700 MeV, and the third one near 2100 MeV. One of these resonances directly corresponds to the investigated N(1710) P$_{11}$.

It is indicative that the model is by itself producing three physical poles for $sols$~4 and 5 where in addition to the two background poles only one (two) bare propagator poles are allowed in the physical region. We say that the model is spontaneously requiring a three-physical-pole solution in spite of the fact that it has not been planned for.

In Table 1. we observe that the  $\chi^2_R$ is rising when we increase the number of resonances. It only reflects the fact that the quality of fit, achieved already for $sol$~4, is not significantly improved by adding new resonances. We are adding them only because we expect that a physical resonance should be produced by the bare propagator pole in the physical region, and not by the interference effect of the nonphysical background poles.

The agreement of the input $T_{\pi N \, \eta N}$  values with the results of the inelastic channel fit for the best three-pole solution ($sol$~6) is given with dotted line in Fig.2, the predictions of $sols$~4 and 5 are not shown because they are very similar to $sol$~3.

The T-matrix values, predicted for the $\pi N$ elastic channel when only \piNetaN channel is fitted, strongly deviate from the input, and are just as an indication given with unlabeled thin grey lines in Fig.1.
\\ \\
\noindent
\emph{Elastic+inelastic channels \dotfill N(1710) P$_{11}$ SURVIVES} \\ \indent
We have simultaneously fitted elastic and inelastic channels and obtained $sols$~7-9.

The overall agreement of the model with the experimental input is achieved using at least three physical poles; one in the vicinity of 1400 MeV (Roper), at least one around 1700 MeV, and the next one near 2100 MeV. The investigated N(1710)P$_{11}$ is needed again, and hence confirmed. However, let us just mention that the 1700 MeV state seems to be degenerated into two near-by poles, but the existing experimental data in \piN and \etaN channels are insufficient to make a firm statement. The  $\chi^2_R$ is falling from $sol$~7 to $sol$~9 indicating that the quality of the fit is being increased by adding new bare propagator poles.

The $sol$~9, the best result of the combined elastic+inelastic channel fit with three bare propagator poles, is as full line shown in Figs~1 and 2 and compared with the the input $T_{\pi N \, \pi N}$ and $T_{\pi N \, \eta N}$.
\\ \\
\noindent
\emph{Other inelastic channels} 

The situation that the interference of background poles produces a physical resonant state, which has occurred when we have fitted the inelastic channel, is repeated for a combined fit in $sol$~9 where the 3-bare-propagator pole solution is generating 4 physical resonances. Therefore, we allow for a 4-bare-propagator pole solution and we obtain $sol$~10, which is with dashed lines shown in Figs.1 and 2. The  $\chi^2_R$ slightly increases with respect to $sol$~9 indicating again that the quality of the fit itself is not improved.

\begin{figure}[!h]
\hspace{-2.5cm}
\includegraphics[width=5cm]{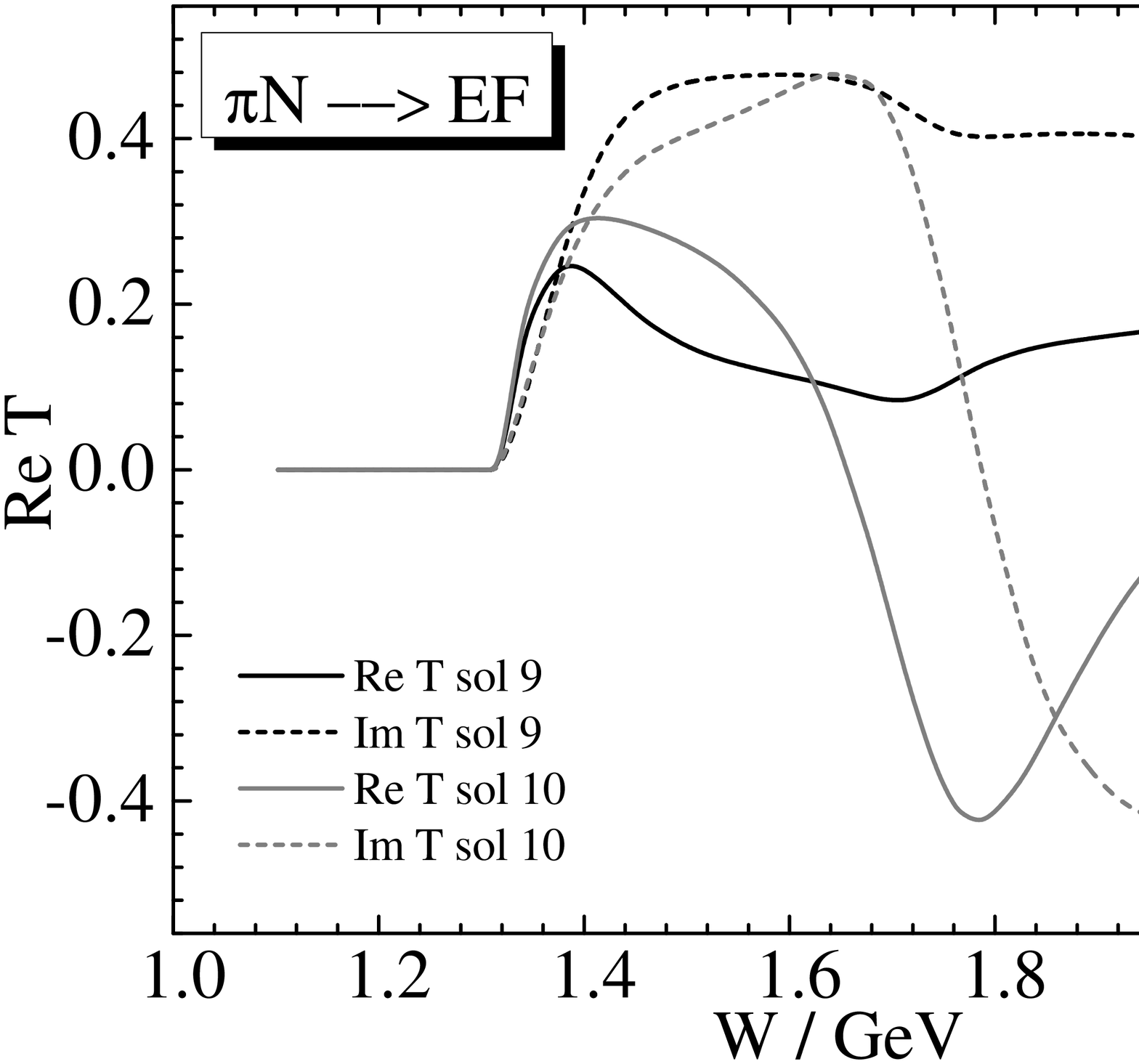}\\
 \caption{The prediction for the $T_{\pi N \, \pi^2 N}$ amplitude for the best three- and the only four-bare-propagator pole solution.}
\label{figure3}
\end{figure}

The obtained 4-pole solution does not significantly differ from $sol$~9 for $\pi N$ and $\eta N$ channels, but we show in Fig.3 that the huge difference between the two solutions appears in the third, effective channel.

Consequently, we do need to measure other inelastic channels (K$\Lambda$ channel for example) in order to distinguish the two and get the unique solution, and that should automatically give us a better insight into the exact structure of all possible P$_{11}$ poles in the 1700 MeV range.

\noindent \\
\textbf{\emph{Summary}} 

We have shown the mechanism how the inelastic data enforce the existence of the N(1710) P$_{11}$ state, our findings coincide with all coupled channel analyses, our model curve is almost identical to the latest the (C-p-$\pi$+) University of Giessen solution \cite{Pen02}, so we claim that the inelastic channels indubitably require at least one P$_{11}$ singularity in the 1700 MeV range.

The PDG confidence rating of the N(1710)~P$_{11}$ should be raised to 4-stars.

Our analysis suggests the presence of yet another resonant state in the 1700 MeV energy range, but its existence (properties) have to be confirmend by including the data from other inelastic channels.

\end{document}